\documentstyle[aps,psfig]{revtex}
\input epsf.tex
\preprint{TIFR/TH 97/26}
\preprint{hep-lat@ftp/9705041}
\def \beq{\begin{equation}}
\def \eeq{\end{equation}}
\begin{document}
\baselineskip=24pt
\begin{center}
\bf{Phase transitions in abelian lattice gauge theories } \\
\vspace{1cm}

\rm{Srinath Cheluvaraja}\footnote[1]{e-mail:srinath@theory.tifr.res.in} \\
Theoretical Physics Group\\
Tata Institute of Fundamental Research \\
Homi Bhabha Road
Mumbai - 400 005, INDIA\\
\end{center}
\vspace{1cm}

\noindent{\bf{Abstract}}\\
We study the phase transition in
the abelian lattice
gauge theory using the Wilson-Polyakov line as the order parameter.
The Wilson-Polyakov line remains very small at strong coupling and becomes
non-zero at weak coupling, signalling a
confinement to deconfinement phase transition.
The decondensation of monopole loops is responsible for this phase transition.
A finite size scaling analysis of the
susceptibility of the Wilson line
gives a ratio for $\gamma/\nu$ which is quite close to
the corresponding value in the 3-d planar
model. 
A scaling behaviour for the monopole loop distribution function 
is also established at the
point of the second order phase transition.
A measurement of the plaquette susceptibility at the transition point shows
that it does not scale with the four dimensional volume as is expected of a
first order bulk transition.
\vspace{0.5cm}
\begin{flushleft}
\end{flushleft}

\newpage
\begin{section}{Introduction}
Lattice gauge theories (LGTs) at non-zero temperature have been the focus
of many investigations in recent years. Their study enables us to
make non-perturbative predictions for the high temperature properties of
gauge theories. Some of the issues which LGTs are expected to clarify are: the
nature of the high temperature phase, the order of the phase transition,
and the relevant elementary excitations at high temperatures.
The pioneering work in
\cite{suss} was the first non-perturbative 
demonstration that quarks are deconfined at high
temperatures. This calculation is done in the strong coupling limit of
the $SU(2)$ LGT. Early
Monte-Carlo simulations \cite{early} provided further support
for the existence of this phase transition.
Since then, there have been many studies of the 
thermodynamic
properties of the $SU(2)$ and the $SU(3)$ LGTs \cite{thmics} which
have provided valuable insights into the high temperature behaviour of
gauge theories.

In this paper, we study the properties of the $U(1)$ LGT at non-zero
temperature. More precisely, we study the $U(1)$ LGT on asymmetric lattices
for which
$N_{\sigma} >>N_{\tau}$. $N_{\sigma}$ is the lattice size in the spatial
direction and $N_{\tau}$ is the lattice size in the temporal direction.
The limit $N_{\sigma}\rightarrow\infty$
and $N_{\tau}$ fixed, with periodic boundary conditions in the temporal 
direction, is the thermodynamic limit for a finite temperature system.
There are several reasons why we have embarked on a study
of this simple model. Firstly, the zero temperature properties 
of this model have been 
inferred
from simulations on large symmetric lattices 
($N_{\sigma}=N_{\tau}$) in \cite{u1x}. In these simulations, 
a phase transition is observed from
a confining phase to a deconfining phase.
Though there was some
controversy about the order of the phase transition, recent simulations on
large lattices strongly suggest that the bulk transition is of first 
order\cite{lang}.
The mechanism of this bulk transition
is quite well understood in
terms of monopole excitations \cite{kogut}.
These monopoles are topological objects
that arise because of the periodicity properties of the action.
The bulk system exists in two phases, a
confining phase in which the monopole currents condense causing complete
Meissner effect,
and a deconfining phase in which the monopoles are too heavy to have any physical
effect.
It would be interesting to see how this picture
of confinement vs deconfinement gets affected at finite temperature.
This is also of relevance from the point of view of some recent claims
of there being a second order phase transition in an extended version of the
$U(1)$ LGT with a monopole chemical potential term \cite{rebbi}.
A definition of a continuum theory is possible at the point of the second order
phase transition, and an investigation of its properties can be quite instructive.

Secondly, the critical  
behaviour of gauge theories at high temperatures 
can be understood
from analogous behaviour of three-dimensional spin models. The strong
coupling analysis in \cite{suss} shows that the partition function of the
$SU(2)$ LGT can be rewritten as a three-dimensional spin model with a
global $Z(2)$ symmetry. 
The deconfining phase
corresponds to the symmetry breaking phase of the spin model,
and the confining phase corresponds to the symmetric phase.
This symmetry arises because finite temperature
gauge systems have an additional global symmetry arising from the periodic
boundary conditions in the temporal direction. 
For systems with a gauge symmetry group $G$, this global symmetry group 
consists of
elements belonging to the center of the gauge group.
Furthermore,
as emphasized in \cite{yaffe},
the order of the phase transition in four dimensions
is expected be dictated by the 
universality classes
present in 3-d spin models having this global symmetry. These
expectations have been borne out for the $SU(2)$ \cite{su2is} 
and the $SU(3)$ LGTs \cite{mont} in which one observes a
second order Ising like and a first order $Z(3)$ like phase transition respectively.
One of our aims is
to examine the same issue in the much simpler $U(1)$ LGT. Unlike the 
non-abelian $SU(N)$ LGTs,
which have a discrete center subgroup, ($Z(N)$), the abelian $U(1)$ LGT has
a continuous center subgroup which is identical to the group itself.
The role of the $U(1)$ group on the confinement to deconfinement transition
is also of some interest from the point of view of the abelian dominance
hypothesis for confinement which holds that a $U(1)$ subgroup controls
the non-perturbative dynamics of non-abelian gauge theories \cite{hoof}.
Finally, as the $U(1)$ LGT already has a bulk phase transition
unlike the $SU(2)$ and $SU(3)$ LGTs, there is the question of
the interplay between this transition with the expected finite temperature
transition. This matter has been recently examined in the context of
$SU(2)$ LGTs using a mixed action; it was found in \cite{gav} that 
the bulk transition and the finite
temperature transition may coincide, making it difficult to distinguish
one from the other. 
Since simulations are always done on finite
lattices, this raises the question whether the earlier studies on symmetric
lattices \cite{creu} were seeing a bulk transition or a 
finite temperature transition. As argued in \cite{gav}, since "zero
temperature" properties are inferred by extrapolating the results on symmetric
lattices, and "finite temperature" properties are inferred by extrapolating the
results on asymmetric lattices, it is not very easy to decide whether one is
seeing a bulk effect or a genuine finite temperature effect. 
A major difference between a bulk transition and a finite temperature
transition, which can in principle distinguish the two transitions,
is the movement of the transition point with the temporal
lattice size. For non-abelian gauge theories, a physically
relevant finite temperature
transition should move towards the weak coupling region in a manner which
is specified by the beta function of an asymptotically free gauge theory.
With these notions in mind, we would like to investigate the finite temperature
properties of abelian lattice gauge theories.

We will mainly consider the Wilson action \cite{wilson} 
for the $U(1)$ LGT, which is
given by
\beq
S=\beta\sum_{n \mu > \nu}\cos (\theta(n\ \mu \nu)) .
\eeq
The $\theta(n\ \mu \nu)$ are the usual oriented plaquette variables:
\beq
\theta(n\ \mu \nu)=\theta(n\ \mu)+\theta(n+\mu\ \nu)-\theta(n+\nu\ \mu)-\theta(n\ \nu).
\eeq
The link variables $\theta(n\ \mu)$ can take values from $-\pi$ to $\pi$.
As mentioned before, the properties of this model at zero temperature are
well known. There is a transition at $\beta \approx 1.0$ that is caused by
a decondensation of monopole currents ( the bulk transition on a $16^4$ lattices
has been located  precisely in \cite{exa} and occurs at $\beta=1.016$).
These monopole currents are defined
on the dual lattice by counting the number of Dirac strings entering or leaving
a three dimensional cube on the original lattice \cite{grand}.
The monopole density on
a link ($\star l$) of the dual lattice is defined as
\beq
\rho(\star l)=\frac{-1}{2\pi}\sum_{p\in c}\bar \theta (p);
\eeq
$\bar \theta (p)$ is extracted from $\theta(p)$ by expressing it as
\beq
\theta(p)=\bar \theta(p)+2\pi n(p),
\eeq
so that it takes values from $-\pi$ to $\pi$.
In the above expression, the monopole current is defined on the
links $\star l$ of the dual lattice which are dual to the cubes $c$ in the
original lattice. As the monopoles form closed loops on the dual
lattice, it is more convenient to measure the total perimeter density
of the monopole current loops. The perimeter density is given by
\beq
\rho=\frac{1}{N_{l}} \sum_{l}\rho(l) \quad ;
\eeq
$N_{l}$ is the total number of links in the lattice.
One observable which is of special relevance to our analysis is 
the Wilson-Polyakov line (henceforth called the Wilson line),
which is defined as
\beq
L(\vec n)=\prod_{n_{0}=0}^{N_{\tau}}\exp(i\theta(\vec n+n_{0}\hat 4\ 
\hat 4))  .
\eeq
This observable is gauge invariant and has played a crucial role in
studies of the deconfinement transition in $SU(N)$ LGTs.
The center symmetry that emerges at finite temperature is the
transformation which multiplies by a constant phase
all the time-like links emanating from
some fixed time slice, namely
\beq
\exp (i\theta(\vec n\ \hat 4)) \rightarrow \exp (i\ \alpha ) 
\exp (i\ \theta(\vec n\ \hat 4)) .
\eeq
Although the action is invariant under this transformation,
the Wilson line transforms as:
\beq
L(\vec n) \rightarrow \exp (i\ \alpha) L(\vec n) .
\eeq
Hence, a non-zero expectation value of the Wilson line signals a
spontaneous breakdown of the center symmetry. The Wilson line is given
the usual physical interpretation by writing it in the form:
\beq
\langle L(\vec n) \rangle =\exp (-\beta F_{q}(\vec n)) .
\label{subt}
\eeq
It measures the free energy ($F_{q}(\vec n)$)
of a static charge in a heat bath
at a temperature $\beta^{-1}$. Hence, a non-zero value of $\langle L(\vec n) \rangle$
indicates
deconfinement of static charges whereas a zero value indicates confinement.
We point out a subtlety in this representation, which has also been noticed
before in the context of the $SU(2)$ LGT \cite{kiss}.
The left hand side is in general
a complex quantity and hence cannot be given a free-energy interpretation.
One way of avoiding this problem is to work with the correlation function
of two Wilson lines which is always a non-negative quantity.
Apart from the Wilson line, we have also studied the susceptibility of
the Wilson line which is defined as
\beq
\chi_{l} = N_{\sigma}^{3} (\langle {\vec s}^2 \rangle - \langle |\vec s| \rangle^{2}).
\quad .
\eeq
The "spin" variable $\vec s$ is constructed in Eq.~\ref{spin}.
The peak in the Wilson line
susceptibility can be used to locate the phase transition.

A related observable that is also relevant is the plaquette susceptibility,
defined as
\beq
\chi_{p}=6 N_{\sigma}^{3} N_{\tau} \sum_{p}(\langle P^{2} \rangle -
{\langle P \rangle}^2)
\quad .
\label{plasusc}
\eeq
In the above equation, $P$ is the average plaquette density in the
system.
The peak in the plaquette susceptibility can also be used to locate
the phase transition.
According to the finite size scaling theory,
the susceptibility of an observable, which blows up 
at a phase transition transition point in the
infinite volume limit,
is expected to scale with the
system size as
\beq
\chi_{p}^{cr} = N^{\alpha} \quad ;
\eeq
here $N$ is the size of the system.
For a first order transition, $\alpha$ equals $d$, the dimensionality of the
system ; for
a second order transition, $\alpha$ equals $\gamma/\nu$, where $\gamma$
and $\nu$ 
are the exponents for the susceptibility and the correlation length
near the transition, which are given by
\beq
\chi = (T-T_{c})^{-\gamma} \ \ \ \
\xi = (T-T_{c})^{-\nu}
\quad .
\label{expo}
\eeq

We will use the above observables to study the phase transition in the
abelian lattice gauge theory.

This paper is organized as follows. Sec 2 contains the
results of our numerical investigations of this model.
In Sec 3 we make some
comments on the mixed action $U(1)$ LGT and compare and contrast it
with that of the of mixed action $SU(2)$ LGT. In Sec 4 we summarize
our conclusions. 
\end{section}

\begin{section}{The $U(1)$ LGT at non-zero temperature.}
In this section we present our numerical analysis of the
model. The system is mimicked at 
finite temperature by
working on an asymmetric lattice ($N_{\sigma} >> N_{\tau}$)
with periodic boundary
conditions in the Euclidean time direction. 
We first briefly describe the numerical procedure that we adopted to obtain
our results. The Metropolis algorithm was used to generate successive
Monte-Carlo configurations. A new link variable $\theta^{\prime}$ was
generated from the old one $\theta$ by adding a number randomly chosen in
the range $(-\alpha,\alpha)$ with uniform probability. The value of $\alpha$
was tuned to get an acceptance of 50 percent. Care was taken so that the link
variables remained in the range $(-\pi,\pi)$. 
Simulations were performed on temporal lattices with
$N_{\tau}=2,3\ 
and\ 4$.
The observables that were measured are: the
monopole density, the Wilson line, the plaquette density, the Wilson line
susceptibility,
and the
plaquette susceptibility.
As the Wilson line is complex ( it is a phase with modulus
equal to one ), we measure its real and imaginary parts separately. If we
simply measure the average value of the real or the imaginary parts, the result
will always be equal to zero because phase transitions
are impossible on finite system
as the tunnelling between the degenerate states
always
restores the symmetry. A rigorous way of studying
symmetry breaking is to study the average value of the Wilson line in the
presence of a small symmetry breaking external field, and then take the
limit of zero field after taking the large volume limit.
A simpler prescription, that is often
employed in studying continuous spin models, is to study the root mean square value
of the order parameter, and this is how we will proceed. The observable
that we have measured as an indicator of spontaneous symmetry breaking
is $\sqrt \langle s \rangle$; $s$ is defined as
\beq
s = {Re L}^{2}+{Im L}^{2}.
\label{spin}
\eeq
Here $Re L$ and $Im L$ refer to the average of the real and imaginary parts
of the Wilson line respectively.
A simple strong coupling analysis (valid for $\beta$ small) 
yields the following effective action
for the Wilson lines:
\beq
S_{eff}=2{(\frac{\beta}{2})}^{N_{\tau}}\sum_{\vec n \vec n^{\prime}}
\cos (\theta (\vec n)-\theta(\vec n^{\prime})) \quad ;
\eeq
$N_{\tau}$ is the temporal extent of the lattice and the $\theta (\vec n)$
variables are the sums of the phases of all the time like links at the
spatial point $\vec n$. This is the action for the three dimensional planar
model which is known to have an order-disorder transition at $\beta_{cr}=0.454$.
For an $N_{\tau}=2$ lattice, this gives the critical coupling to be
approximately $0.95$. Thus we expect our lattice model to have a phase
transition at $\beta \approx 0.9$. This strong coupling argument is valid
only if the phase transition takes place in the strong coupling regime.
Nevertheless, the strong coupling approximation provides a simple way of
seeing how a three-dimensional spin model emerges from the four dimensional
gauge theory.

A more direct way of seeing the appearance of an effective 3-d spin model,
without using a strong coupling approximation,
is to use the dual representation of the 4-d $U(1)$ LGT.
The dual representation
is given by
\beq
Z\ =\ \sum_{m_{\mu}(r)}\exp -\frac{1}{2\beta}\sum_{n\ \mu \nu}
(\partial_{\mu}\phi_{\nu}-\partial_{\nu}\phi_{\mu})^{2}
+2\pi i\sum_{n}m_{\mu}(n)\phi_{\mu}(n)
\eeq
This describes a gas of closed monopole loops ($m_{\mu}(r)$) which interact
by a 4-dim Coulomb potential. This system has a phase transition
which takes place as a result of the competition between the energy and
the entropy of the monopole loops.
The fields $\phi_{\mu}(r)$ can be integrated out to give
\beq
Z\ =\ \sum_{m_{\mu}(r)}\exp (-2\pi^{2}\beta\sum_{n\ n^{\prime}}
m_{\mu}(n)G_{\mu\ \nu}(n-n^{\prime})m_{\nu}(n^{\prime}))
\eeq
At non-zero temperature, the time direction is finite, and the Green's function
satisfy periodic boundary conditions. The four dimensional Green's function can
be rewritten as:
\beq
G_{\mu\ \nu}(r-r^{\prime}) \approx T \tilde G_{\mu\ \nu}(\vec r-\vec r^{\prime})
\quad ;
\eeq
here $\tilde G_{\mu\ \nu}(\vec r-\vec r^{\prime})$ is the 
3-dimensional Green's function.
As the gas is finite in one direction, one again has a gas of monopole
loops which now effectively interact with a 3-dim Coulomb interaction.
From the point of view of the effective  three-dimensional 
planar model, the monopole loops behave like vortex lines.
The entropy of large loops in three-dimensions 
is smaller than in the four-dimensional case
and this shifts the transition.
The order of the transition cannot be deduced from these
energy arguments and has to be determined 
by doing a finite size scaling analysis.
We show in Fig.~\ref{order} the 
variation of $\sqrt \langle s \rangle$ with $\beta$ on a $6^3\ 2$ lattice. The observable
$\sqrt \langle s(\vec n) \rangle$ is close to zero at small $\beta$
and rises smoothly across the critical
value. The $U(1)$ monopole density variation is shown in
Fig.~\ref{monopole}.
There is a fall in the monopole density across the transition which
coincides with the rise in the order parameter. In both cases, the variation
is quite gradual and we would suspect that we are in the vicinity of a
second order transition. This is made further suggestive by the gradual
rise in the plaquette expectation value (see Fig.~\ref{plaqu}).
To determine the order of the transition, we
perform a finite size scaling analysis of the susceptibility of the 
order parameter ($\chi_{l}$).
We have done the finite size scaling study on lattices of temporal extent 
$N_{\tau}=3$ and $N_{\tau}=4$.
For this method to work, it is crucial 
that we
are very close to the pseudo-critical point corresponding to the
lattice that we are working on.
The histogram method is
used to extrapolate observables from one value to a nearby 
neighbouring value. The pseudo-critical
point is located in this way by looking at the peak in the susceptibility
of the order parameter.
The behaviour of the susceptibility near the transition 
on $6,9,12$ and $16$ sized spatial
lattices (keeping the temporal extent fixed at $N_{\tau}=3$)
is shown in Fig.~\ref{susc}.
These results were got after 200,000 measurements
on $N_{\sigma}=6,9$ lattices while 150,000 measurements were made on
the $N_{\sigma}=12$ lattice and 100,000 measurements were made on the
$N_{\sigma}=16$ lattice. All the measurements were made after ignoring
the first 50,000 Monte-Carlo iterations. The errors were estimated by
binning the data.
The finite size scaling theory predicts
\beq
\chi \approx N_{\sigma}^{d}
\eeq
for a first order phase transition; d being the effective dimension of the
system (in this case $d=3$); for a second order transition, the
prediction is
\beq
\chi \approx N_{\sigma}^{\frac{\gamma}{\nu}};
\eeq
$\gamma$ and $\nu$ are the exponents in Eq.~\ref{expo}.
The peaks in the susceptiblity were 
fit to an $N_{\sigma}^{\alpha}$ dependence
and a good fit is obtained (Fig.~\ref{strline}).
From our fit we determine $\gamma/\nu$ to be 1.95 with an error of 0.13.
For the three dimensional planar model, which is the effective spin model
with which we would like to compare our results, the ratio $\gamma /\nu=1.97$.

We have performed another test for the order of the phase transition by
studying the distribution of the monopole loops. Since the monopole loops,
which now behave as vortex lines,
are responsible for driving the phase transition, they should also exhibit
a scaling behaviour at the point where the transition becomes second order.
At the point of the second order phase transition, there are fluctuations on all
length scales and the monopole loops come in all sizes and shapes.
At the transition point, 
the monopole loop distribution function, $p(l)$, should scale as
\beq
p(l)=\frac{1}{l^{\tau}}
\eeq
with an exponent $\tau$ which is independent
of the volume. The function $p(l)$ is defined as the probability of
finding a loop of length $l$ at a site. 
We have calculated $p(l)$
for various lattices at their pseudo-critical points and we find that
$p(l)$ can be nicely fitted to the above form. 
The value of $\tau$ that we get (using data on a $N_{\sigma}=16$ lattice)
is $2.27\pm.003$. Unfortunately, we are not able to
compare it with a theoretical calculation. Nevertheless, it is interesting
that the topological excitations, in this case the monopoles, 
exhibit a scaling dependence at the transition point.
Fig.~\ref{monscale}
shows $p(l)$ for loops of length $l$ upto 14 on different lattices. The plot
shows that they fall on a straight line independent of the volume. The slight
distortions for the $N_{\sigma}=6$ lattice are due to finite size corrections
when the loop length is larger than the lattice size. A study of larger loop
lengths can also be made but the algorithm that we have used to count the loops
slows down when the loop lengths become very large, and so we have not
studied larger loops.
Away from the critical point, $p(l)$ either falls rapidly to zero
(in the deconfining phase)
or remains constant (in the confining phase). A similar study of the monopole
loop distribution function was made in \cite{grady} for the abelian projected
monopoles in the $SU(2)$ theory near the continuum limit.

Now we say a few words on the location of this transition as compared to
the location of the bulk transition.
The transition point (which is located by looking at the peak in the
susceptibility of the order parameter) shifts as a function of $N_{\tau}$ as:

$N_{\tau}$ \hspace{2cm} 2 \hspace{2cm} 3 \hspace{2cm} 4

$\beta_{cr}$ \hspace{2cm} 0.9297(3) \hspace{0.8cm} 1.012(2) \hspace{0.9cm} 1.032(2)

\noindent
The above critical values are those obtained on $N_{\sigma}$
of $16,16$ and $12$ respectively.
The bulk transition on a $16^4$ lattice was located at $\beta=1.016$ in
\cite{lang}.
Since the bulk transition is known to be of first order \cite{lang},
and the
transition that we have observed is of second order, we are observing a
change in the order of the phase transition as a function of $N_{\tau}$.
The nature of the transition that we have observed 
can be further studied by monitoring the behaviour of the
plaquette susceptibility which was defined in Eq.~\ref{plasusc}.
The behaviour of the plaquette susceptibility 
for a $6^3\ 3$ and a $12^3\ 3$ lattice is
shown in Fig.~\ref{plasus}.
This graph clearly shows that the plaquette susceptibility does not
scale with the four dimensional volume as is expected of a
first order bulk transition.

The above observations show that the transition that we have observed 
(on an $N_{\tau}=3$  lattice) is
a deconfinement transition whose scaling behaviour is
quite distinct from that of the
bulk transition. The order of the transition as seen by the Wilson line is
a second order transition with the ratio $\gamma/\nu$ which has 
approximately the same
value as that in the three-dimensional planar model.
Since the plaquette susceptibility does not scale as a first order
bulk transition, we are not observing the bulk transition at the point
where we are observing the finite temperature transition.
The finite size scaling analysis can be repeated on an $N_{\tau}=4$ lattice
and the peak in the susceptibility  of the order parameter
as a function of volume is shown in
Fig.~\ref{susc2}. This scaling again suggests a second order transition with
a value for $\gamma/\nu$ which is again in good agreement with that of the 
three-dimensional planar model.

We now place our results in the context of some recent developments.
The studies in \cite{gav} considered the finite temperature properties
of the mixed action $SU(2)$ LGT which is defined by
\beq
S = (\beta_{f}/2) \sum_{p}tr_{f}\ U (p)+(\beta_{a}/3) 
\sum_{p}tr_{a}U(p).
\eeq
This model is known to have lines of first order bulk transitions 
in the $\beta_{f},\beta_{a}$
plane \cite{creu}.
It was found in \cite{gav} that the deconfinement transition of the pure $SU(2)$ LGT,
which is of second order,
continues into the $\beta_{f}$,$\beta_{a}$ plane and joins the
line of first order bulk transitions.
Based on this observation in \cite{gav}, a possibility was considered
where the there is either only a bulk or a finite temperature transition in
the $SU(2)$ LGT. 
On the other hand, there may also be a very small but finite
separation in the two transitions which cannot be resolved on the
lattice sizes used in the simulations.
In our case, the relevant coupling space is the $\beta$, $T$ plane. 
Our simulations of the $U(1)$ LGT show that there is a deconfinement
transition which is of second order and that the
plaquette susceptibility does not scale as a first order bulk transition
at the transition point.
The transition is shown to shift from its bulk value 
as a function of $N_{\tau}$ but this shift
is very small, though still discernible.
We also place our results in the light of the expectations of
Svetitsky and Yaffe \cite{yaffe}. According to their general arguments, 
the critical begaviour of the 
$U(1)$ LGT at finite temperature is expected to fall in the same
universality class as that of the 3-d planar model, which is known to have a
second order phase transition. Our finite size scaling analysis
on the $N_{\tau}=3$ lattice definitely rules out a first order 
phase transition
and indicates that the ratio $\gamma/\nu$ has approximately the same value
as in the 3-d planar model. 

We conclude this section with a proposal for the finite temperature
phase diagram for the U(1) LGT. In the bulk system (which corresponds to
taking the limit $N_{\sigma}=N_{\tau} $ and
$N_{\sigma}\rightarrow \infty$),
there is the
monopole driven phase transition. 
On asymmetric lattices of very small
temporal extent, we expect the bulk system to look like a three
dimensional $U(1)$ gauge theory. The three-dimensional $U(1)$ LGT
has no bulk transition.
On asymmetric lattices of very large
temporal extent we expect to see the bulk transition. Lattices of
intermediate temporal extent will exhibit a complicated crossover from a
four-dimensional bulk system to a three-dimensional bulk system.
Our simulations on an $N_{\tau}=3$ lattice show that there is a second
order phase transition
which has the same value of $\gamma/\nu$ as in the three-dimensional
planar model. This transition is a deconfinement transition as is indicated by
the behaviour of the fundamental Wilson line.
However, there is no bulk transition on this lattice.

\end{section}

\begin{section}{Mixed action $U(1)$ LGT.}
Since lattice actions are anyway not unique, we can always construct
more complicated looking lattice actions and examine their properties.
A simple generalization of the action in Sec 1 is the mixed action which
is defined by
\beq
S = \beta_{1}\sum_{p}\cos (\theta(p))+\beta_{2} \sum_{p}\cos (2\theta (p)).
\eeq
The two pieces of the above action are different 
from each other only
insofar as their periodicity properties are concerned. In the naive
continuum limit, $a\rightarrow 0$, the second term is like an irrelevant
coupling and is not expected to change the long distance properties of the
theory.
The zero temperature properties of this action have been studied 
in \cite{mixu1}
and it has a rich phase structure of first and second order transitions.
This system can also be studied at finite temperature just as we
did for the $\beta_{2}=0$ theory.
Again, from a simple strong coupling analysis of the mixed action $U(1)$
LGT,
we get an effective theory
of spins which is that of the mixed planar model
\beq
S_{eff} =2(\frac{\beta_{1}}{2})^{N_{\tau}}\sum_{\vec n \vec n^{\prime}}
\cos (\theta(\vec n)-\theta(\vec n^{\prime})) + 2(\frac{\beta_{2}}{2})^{N_{\tau}}
\sum_{\vec n \vec n^{\prime}}\cos (2\theta(\vec n)-2\theta(\vec n
^{\prime})).
\eeq
Putting $\beta_{1}=0$ gives a three-dimensional planar model
of spins, the only difference being in the periodicity properties of
the action.
Hence, the finite temperature properties of the mixed model in the
$\beta_{1}=0$ limit should be identical to those in the $\beta_{2}=0$ limit.
In particular, the previous statements 
regarding the order of the transition will
also be true in this limit.
A surprising feature of the mixed planar model is that it posesses a
region of first order phase transitions for some values of $\beta_{2}$
\cite{klei}. This implies a similar region of first order transitions
in the mixed $U(1)$ LGT for a segment of $\beta_{2}$ values. The order of
the finite temperature transition changing in the direction of an irrelevant
coupling has also been discussed in the context of the mixed action $SU(2)$
LGT \cite{gav}.

From the above analysis, it is clear that there are many similarities
between the $U(1)$ LGT and the mixed action $SU(2)$ LGT. We now show that
the similarity also extends to the construction of the relevant order
parameters in the two models.
The mixed action $SU(2)$ LGT \cite{creu}
is defined by
\beq
S =\frac{\beta_{f}}{2}\sum_{p}tr_{f}U(p) +\frac{\beta_{a}}{3}\sum_{p}
tr_{a} U(p); 
\eeq
$tr_{f}$ and $tr_{a}$ denote the traces in the fundamental and
the adjoint representations respectively. The limit $\beta_{a}=0$ describes
an $SU(2)$ LGT and the limit $\beta_{f}=0$ describes an $SO(3)$ LGT.
The order parameter of the finite temperature transition in the $SU(2)$ LGT
is the Wilson line in the fundamental representation which is defined as
\beq
L_{f}(\vec n)=Tr_{f} \prod_{n_{0}=0}^{N_{\tau}}U(\vec n+n_{0}\hat 4\ \hat 4).
\eeq
In the $SO(3)$ LGT ($\beta_{f}=0$), this observable is identically
zero because of the following $local$ $Z(2)$ symmetry:
\beq
U(\vec n\ \hat 4) \rightarrow Z(\vec n) U(\vec n\ \hat 4);
\label{local}
\eeq
$Z(n)$ can take the values $+1$ or $-1$ at any site.
For the $SO(3)$ LGT, the appropriate order parameter is the
Wilson line in the adjoint representation,
\beq
L_{a}(\vec n)=Tr_{a} \prod_{n_{0}=0}^{N_{\tau}}U(\vec n+n_{0}\hat 4\ \hat 4)
\quad ,
\eeq
which is invariant under the local $Z(2)$ transformation in Eq.~\ref{local}. 
For the group $SU(2)$, $L_{f}$ and $L_{a}$ are related by
\beq
L_{a}(\vec n) = {L_{f}(\vec n)}^{2} -1.0 \quad .
\eeq
In the mixed action $U(1)$ LGT we are faced with a similar problem
in defining the order parameter for the $\beta_{1}=0$ theory. In this
limit, the Wilson line defined in Sec 1 is identically zero because of
the following local symmetry:
\beq
\exp (i\theta(\vec n\ \hat 4)) \rightarrow Z(\vec n)
\exp (i\ \theta(\vec n\ \hat 4)).
\eeq
The correct order parameter to use in this limit is
\beq
L_{2}(\vec n)=\prod_{n_{0}=0}^{N_{\tau}}\exp(i\ 2\theta(\vec n+n_{0}\hat 4
\ \hat 4))
\quad ,
\eeq
which is analogous to the Wilson line in the adjoint representation
of $SU(2)$. The relationship between $L_{2}(\vec n)$ and $L(\vec n)$
is $L_{2}(\vec n)={L(\vec n)}^{2}$. 

Though there are similarities 
at the formal level, and in the phase
structure
of the mixed action abelian and mixed action
non-abelian theories, there are, of-course, many 
important
differences between the two systems.
An important difference is that, unlike in the abelian LGT, the
critical coupling
in the $SU(2)$ LGT is expected to
scale with $N_{\tau}$ according to the beta function of the Yang-Mills
theory.
There is strong evidence for this asymptotic scaling from simulations
on very large temporal lattices \cite{karsch}. 

The purpose of this section was only to indicate that many of the issues
such as the mixing of the bulk and finite temperature transitions which
have been raised recently can all be explored in the much simpler mixed
action abelian lattice gauge theory. Also, because the 
physical properties of this model
are well
understood in terms of the monopole excitations, this model
may prove useful
in investigating these issues.
\end{section}
\begin{section}{Conclusions}
In this paper we studied the 
$U(1)$ LGT using the Wilson line as the order
parameter. We have found that there is a transition into a deconfining phase
at large coupling which is driven by the decondensation of monopole loops.
The monopole loops, however, effectively interact like the vortices in the
three-dimensional planar model.
The deconfining phase breaks the global $U(1)$
symmetry present in the theory. A finite
size scaling analysis of the susceptibility of
the Wilson line indicates that the transition is of second order 
( on lattices of temporal size $N_{\tau}=3$ and $N_{\tau}=4$) with a 
ratio for $\gamma/\nu$ which has almost the same value as in the 
three-dimensional planar model. A scaling form for the
distribution of the monopole loops 
was also established at the point of the second order
phase transition. This transition was also examined by studying the
plaquette susceptibility at the transition point. The plaquette
susceptibility (on a  $N_{\tau}=3$ lattice)
does not scale as is expected of a first order bulk transition.
There is also a small shift in the transition point from the bulk value.
We have also pointed out that 
many of the recent issues concerning the
mixing of the bulk and finite temperature transitions can also be raised in the
abelian lattice gauge theory. Since the abelian theories
are well understood in terms of their monopole excitations, some of these
issues can perhaps be clarified.
\end{section}

\newpage
\begin{figure}[ht]
\caption {Variation of the order parameter on a $6^3\ 2$ lattice.}
\label{order}
\end{figure}
\begin{figure}[ht]
\caption {Plaquette expectation value on a $6^3\ 2$ lattice.}
\label{plaqu}
\end{figure}
\begin{figure}[ht]
\caption{Monopole density on a $6^3\ 2$ lattice.}
\label{monopole}
\end{figure}
\begin{figure}[ht]
\caption {
Susceptibility of the order parameter 
near the transition on $6,9,12$ and $16$ size
spatial lattices.}
\label{susc}
\end{figure}
\begin{figure}[ht]
\caption {A fit of the maximum value of the logarithm of the
susceptibility to $ln N$.}
\label{strline}
\end{figure}
\begin{figure}[ht]
\caption {Scaling of the monopole loop distribution function 
$p(l)$ with
loop length (l). The graph shows the logarithm of $p(l)$ as a function
of the logarithm of $l$.
}
\label{monscale}
\end{figure}
\begin{figure}
\caption{ A finite size study of the susceptibility of the order
parameter on a $N_{\tau}=4$ lattice.}
\label{susc2}
\end{figure}
\begin{figure}
\caption{ The scaling of the plaquette susceptibility near the
phase transition. This is shown on $6^3\ 3$ and $12^3\ 3$ lattices.
}
\label{plasus}
\end{figure}
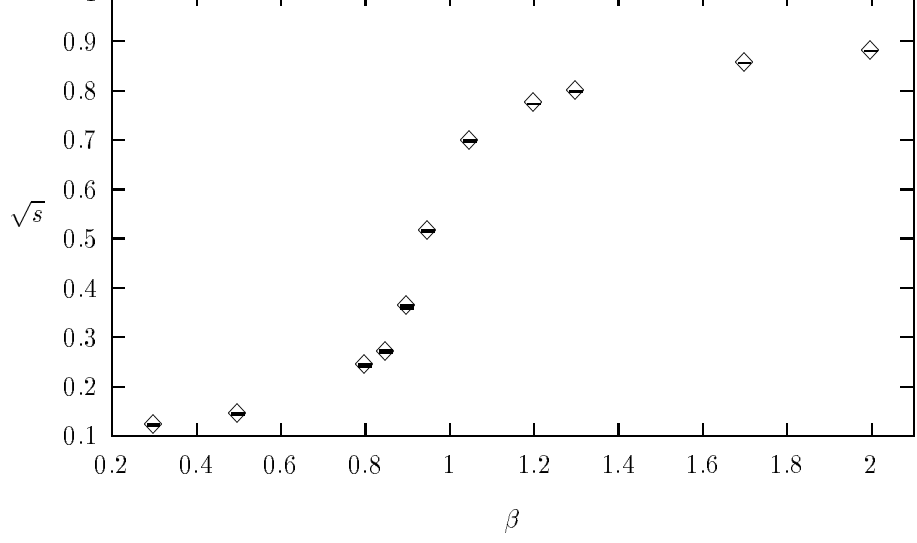
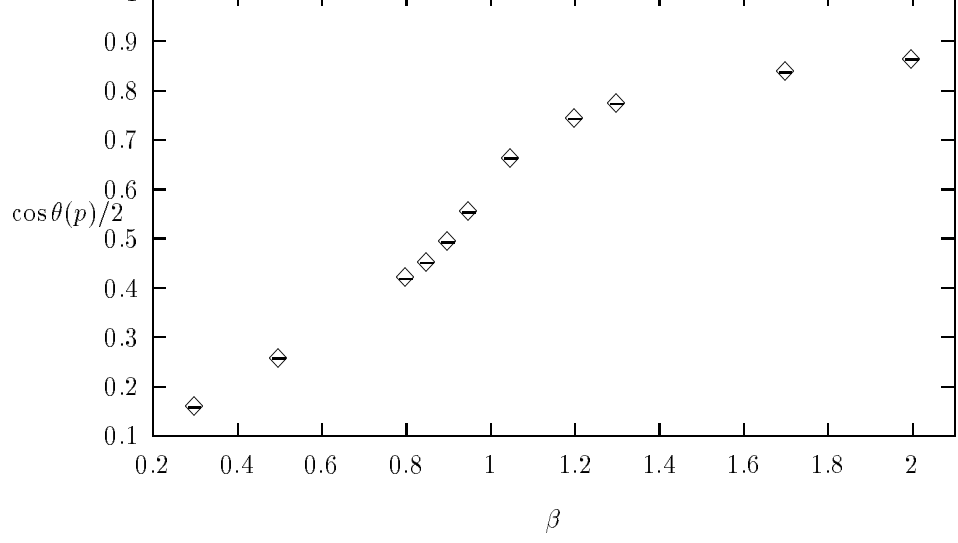
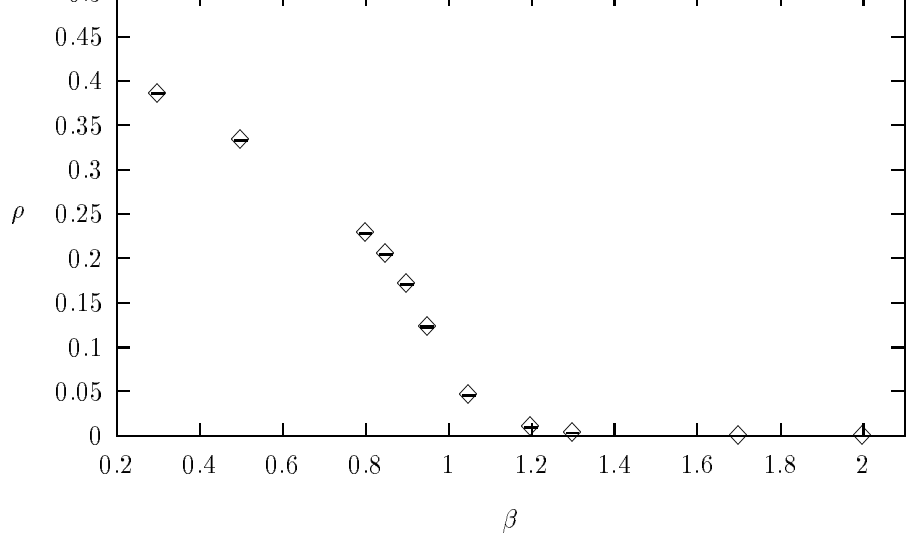
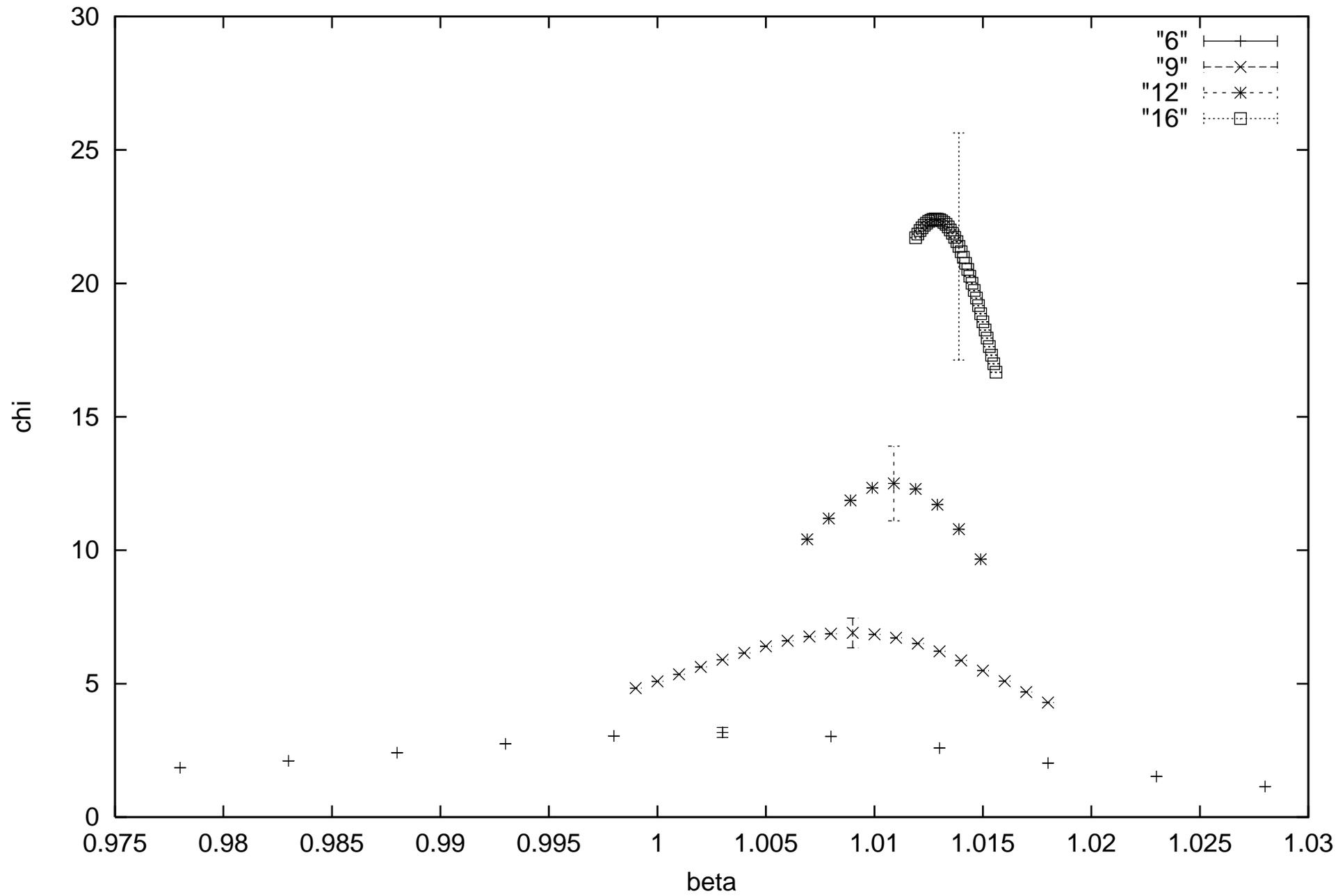
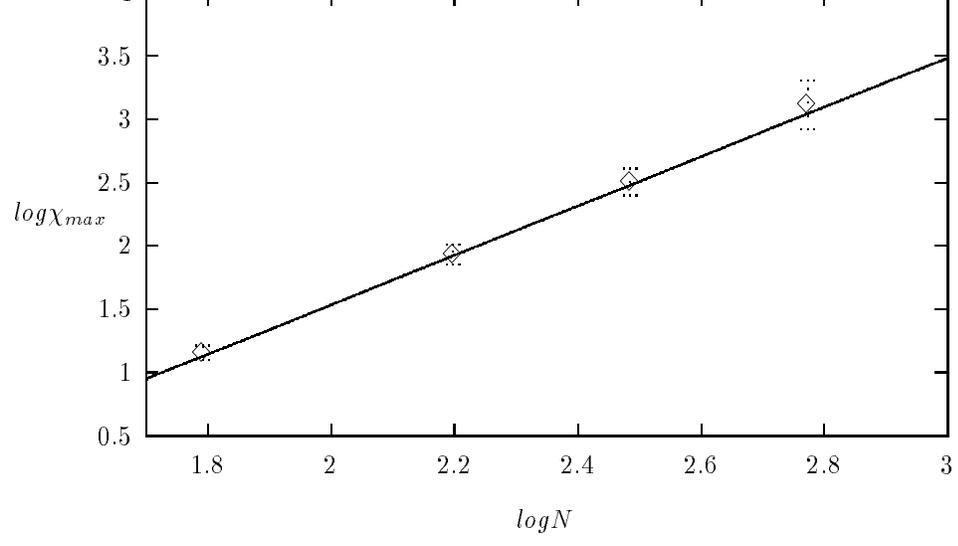
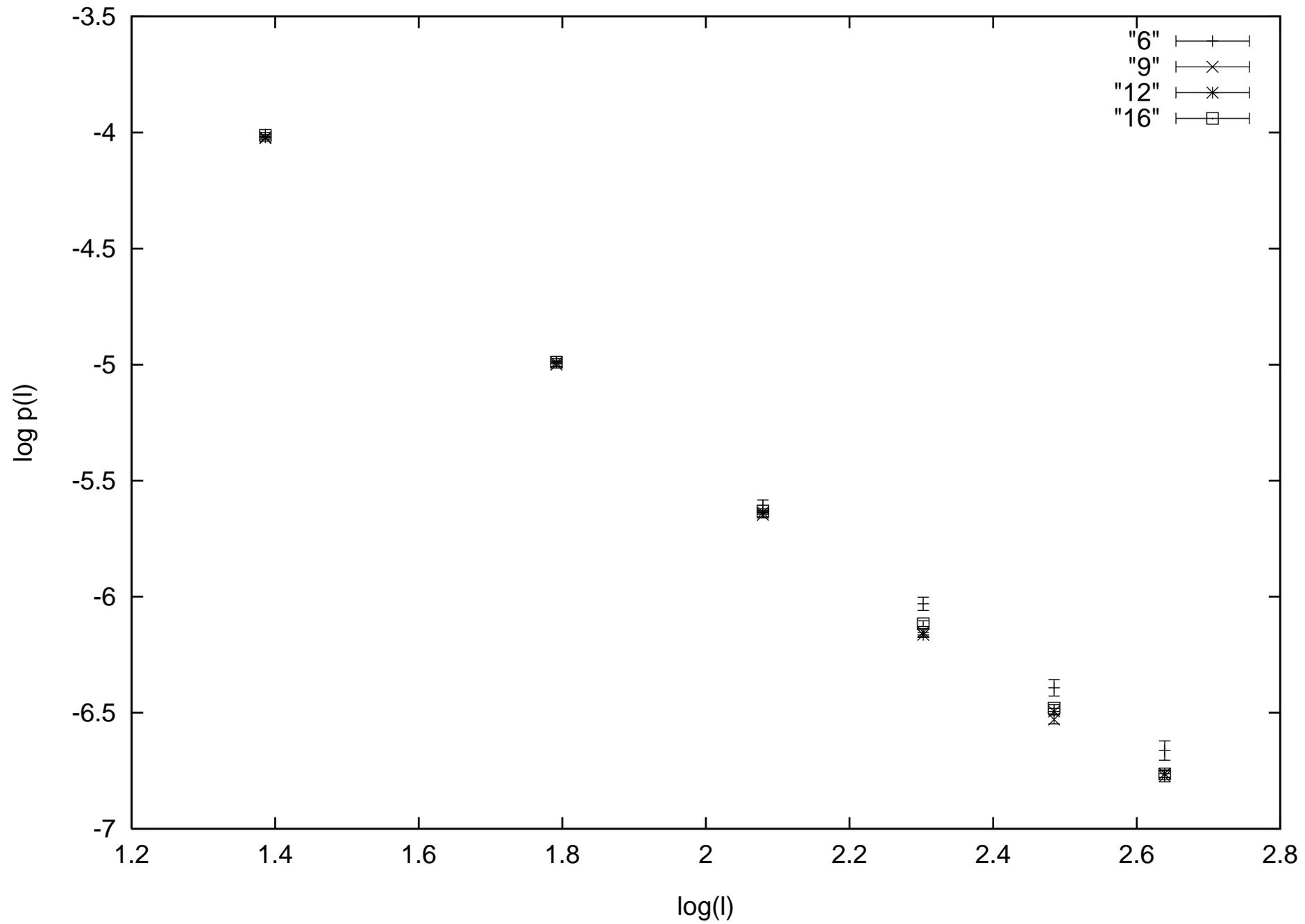
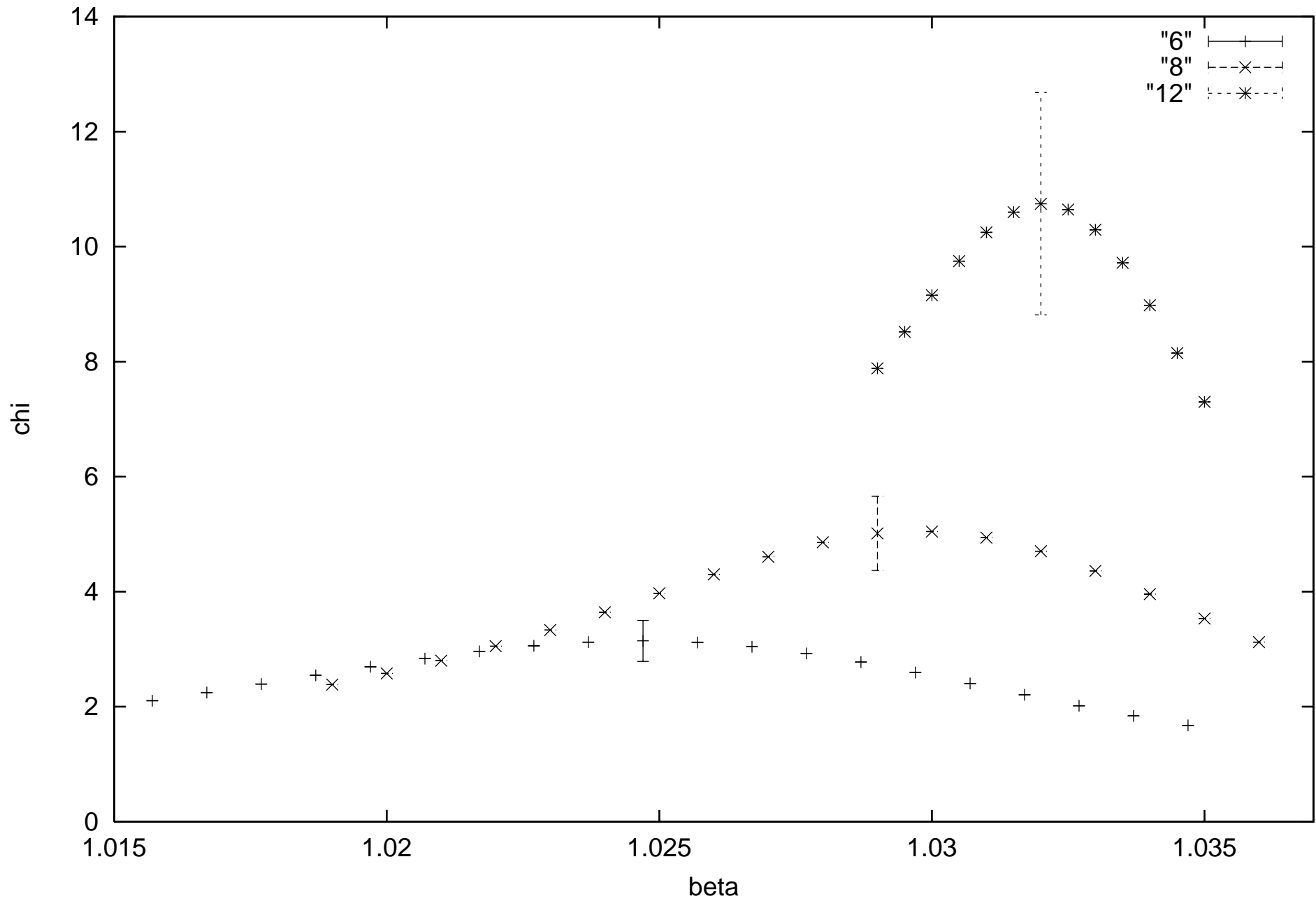
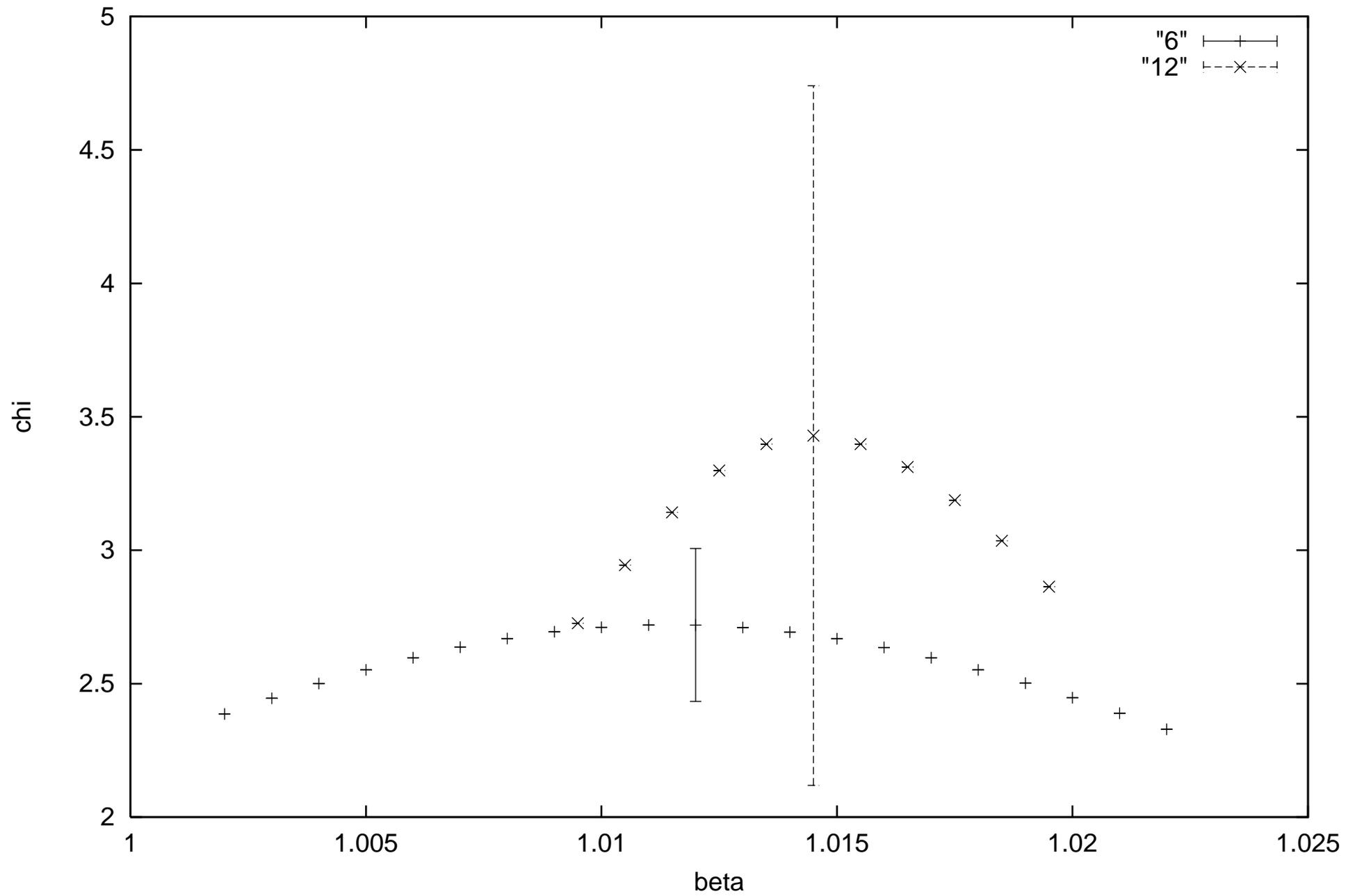
\end{document}